\documentstyle[preprint,eqsecnum,aps]{revtex}
\begin{document}
\draft
\tighten

\title{\Large\bf Deep Inelastic Structure Functions in Light-Front QCD: 
	Radiative Corrections}

\author{{\bf A. Harindranath$^a$, Rajen Kundu$^a$, and Wei-Min Zhang$^b$} \\
$^a$Saha Institute of Nuclear Physics, 1/AF, Bidhan Nagar, 
	Calcutta 700064 India \\
$^b$Institute of Physics, Academia Sinica, Taipei, Taiwan 11529, ROC}

\date{May 12, 1998}

\maketitle

\begin{abstract}
Recently, we have introduced a unified theory to deal with perturbative 
and non-perturbative QCD contributions to hadronic structure functions 
in deep inelastic scattering. This formulation is realized by combining 
the coordinate space approach based on light-front current algebra 
techniques and the momentum space approach based on Fock space expansion 
methods in the Hamiltonian formalism of light-front field theory. 
In this work we show how a perturbative analysis in the light-front 
Hamiltonian formalism  leads to the factorization scheme we have proposed 
recently. The analysis also shows that the scaling violations due to 
perturbative QCD corrections can be rather easily addressed in this 
framework by simply replacing the hadron target by dressed parton target 
and then carrying out a systematic expansion in the coupling constant 
$\alpha_s$ based on the perturbative QCD expansion of the dressed parton 
target. The tools employed for this calculation are those 
available from light-front old-fashioned perturbation theory. We present 
the complete set of calculations of unpolarized and polarized deep 
inelastic structure functions to order $\alpha_s$. We extract the 
relevant splitting functions in all the cases. We explicitly verify all 
the sum rules to order $\alpha_s$. We demonstrate the validity of 
approximations made in the derivation of the new factorization scheme.  
This is achieved with the help of detailed calculations of the evolution 
of structure function of a composite system carried out using multi-parton 
wavefunctions.
 \end{abstract}

\vspace{0.5in}


\section{Introduction}
A general cross section in hadron physics contains both short distance and
long distance behavior and hence is not accessible to perturbative QCD.
Factorization theorems \cite{pQCD} allow one to separate the two behaviors in a
systematic fashion. Usually the short distance (perturbative) properties 
are calculated with Feynman diagrams where the most popular choice of 
regulator is dimensional regularization. So far there is no non-perturbative
implementation of dimensional regularization in field theory. The long distance
(non-perturbative) part is given in terms of a set of operator matrix
elements which are left for computation, for example, in lattice gauge
theory. Since factorization scale is an artifact reflecting our present
inability to do computations in QCD, the two  sectors (perturbative and
non-perturbative) should merge smoothly. Since currently available 
formalisms employed to tackle the two sectors use different
regulators, different degrees of freedom, etc., this goal is difficult to
accomplish in practice. It is desirable to have a method of calculation
where same formalism is used to deal with both perturbative and
non-perturbative regions of QCD.    

Recently we have proposed a new method \cite{paper1} of calculation of deep
inelastic structure functions that combine current algebra techniques and
Fock space expansion methods in light-front field theory in a Hamiltonian
framework. We have arrived at expressions for various structure functions 
as the Fourier transform of hadron matrix elements of different components 
of bilocal 
vector and axial vector currents on the light-front. By expanding the state
of the hadron in terms of multi-parton wave functions, non-perturbative QCD
dynamics underlying the structure functions can be explored. We have also
presented a novel analysis of power corrections based on light-front power
counting. 

In this work we show that a perturbative analysis in the light-front 
Hamiltonian formalism  leads to the factorization scheme proposed in 
Ref. \cite{paper1}. The analysis also shows that
the scaling violations due to perturbative QCD
corrections can be rather easily addressed in this framework by simply
replacing the hadron target by dressed parton target and then carrying 
out a systematic expansion in the coupling constant $\alpha_s$ based on 
the perturbative QCD expansion of the dressed parton target. The tools 
used are those belonging to light-front old fashioned perturbation 
theory \cite{Brodsky,Zhang93,Hari96} which employs transition matrix 
elements and energy denominators instead of Feynman propagators. 

The advantage of using the light-front Hamiltonian formulation is that
the matrix elements can be naturally defined in the light-front gauge 
($A^+=0$ in light-front coordinates). With this gauge choice, there is 
no need for the path ordered exponential between fermion field operators 
in the bilocal current which is mandatory in other popular gauge 
choices.  Since we do not have to deal with four dimensional integrals 
involving Feynman propagators we do not encounter some of the problems 
associated with the use of the non-covariant gauge condition $A^+=0$ in 
usual covariant perturbative QCD calculations.  

Meanwhile, the evaluation of the matrix elements in our approach (see 
sections IV and V) is straightforward, which also greatly clarify the 
physical picture of DIS. For example,  matrix elements of the transverse 
component of the bilocal vector and  axial vector currents are easy to 
analyze in the present method. In sharp contrast, there are well-known 
problems\cite{Van93}  associated with $\gamma_5$ in dimensional  
regularization. Also  presence of quark masses poses no problem
in our calculations which has been bothersome for a long time in the
standard OPE or Feynman diagram approach of DIS.

Since our approach deals with probability amplitudes,
real and virtual processes are calculated to the same order without any
difficulties,  in contrast with the Altarelli-Parisi method\cite{AP} 
which deals with probability densities. We first present result for the 
unpolarized structure functions $F_2(x)$ for 
dressed quark (extracted from the plus component of the vector bilocal)
and gluon target to order $\alpha_{s}$. We extract the relevant splitting
functions. In these cases we also explicitly verify the longitudinal
momentum sum rule. We also present result for the polarized structure
function $g_1$ for a dressed quark. Furthermore, interference effects 
are straightforward to handle. As a result, we explicitly show the 
invalidity of the popular twist classification \cite{Jaffe}. We 
demonstrate this by analyzing the matrix element of the transverse 
component of the bilocal vector and axial vector currents (see Secs. IV 
and V for details). In the conventional OPE analysis \cite{Shu82}, it is 
customary to ignore the non-trivial structure of the state, and consider 
the operator structure alone to draw conclusions. 
A case in point is the transverse component of the bilocal vector current.
This operator is not diagonal and in the twist analysis of Ref.
{\cite{Jaffe}} will appear to have twist three. Hence it appears, as if this
operator matrix element cannot have parton interpretation. However our
explicit calculations show that this operator matrix {\it matrix element} is
indeed twist two and has the familiar parton interpretation. This becomes
clear only {\it after} the evaluation of the matrix elements
 which includes off-diagonal ones \cite{Hari97}. 

For a second example, in the case of the transverse
polarized structure function, it is popular to ignore quark mass
and stress quark-gluon
correlations. We show by explicit calculations that the operators that
involve $ \gamma_5$ which do not have
explicit dependence on the quark mass $m_q$, come 
out proportional to $m_q$  when matrix element is
taken between dressed quark states.  In particular all the operators 
contribute at the same level to the evolution of $g_T$. This is shown to be
essential\cite{Hari97b} for the $g_2$ structure function to obey 
Burkhardt-Cottingham  sum rule \cite{BC}
in perturbative QCD.

In this work we also provide a detailed calculation of the evolution of  
$F_2$
structure function for a hadronic bound state. Entire analysis is carried
out using multi-parton wave functions in momentum space. Both real and
virtual processes are accounted for and in the lowest order analysis one
begins to see the emergence of the standard evolution equation.  
The detailed analysis justifies the approximations made in Sec. II which
lead to factorization to all orders in perturbation theory. Then 
what remains unsolved is the nonperturbative contribution that is defined in 
the same framework. At present, the  nonperturbative QCD dynamics and hadronic 
bound states on the light-front are indeed explored in this same framework.
Therefore, we hope that the present work (ref.\cite{paper1} and the present 
paper) can really provide a natural connection of the fully theoretical
understanding to the experimental phenomena of DIS. Note that for exclusive
processes, factorization has been proven using light-front formalism by
Brodsky and Lepage \cite{Brodsky}.

The plan of this paper is as follows. In section II we present a  
perturbative analysis in the light-front 
Hamiltonian
formalism which leads to the factorization and to the concept of the structure
function of a dressed parton in DIS. The tools necessary for the evaluation 
of these
functions are discussed in Sec. III. Unpolarized and polarized dressed 
parton distribution functions are discussed in Secs. IV and V
respectively. We also present the explicit verification of the appropriate
sum rules in these sections. A detailed analysis 
of the structure function of bound states is carried out in Sec. VI which
justifies a posteriori the approximations made in the study in Sec. II.
Finally Sec.  VII contains discussion and conclusions.  

\section{Factorization: A Perturbative Analysis}
In this section we show in detail, how the factorization picture
discussed in Ref. \cite{paper1} emerges in a perturbative analysis carried
over to all orders in the case where the bilocal operator involved
does not change the
particle number. The analysis leads to the concept of the structure function
of a dressed parton in DIS.

To explicitly demonstrate the factorization picture on the light-front,
we consider the $F_2$ structure function as a specific example in
this section. For simplicity we drop reference to the flavor, then
\begin{eqnarray}
{F_2(x,Q^2) \over x} && = { 1 \over 4 \pi} \int d \xi^- e^{-{ i \over 2} P^+
\xi^- x}  \langle PS  \mid
\Big [ (\psi^+)^\dagger(\xi^-) \psi^+(0) - 
(\psi^+)^\dagger(0) \psi^+
(\xi^-) \Big ] \mid PS \rangle. \nonumber \\
&& 
\end{eqnarray}
From the discussion in Sec. V.D of \cite{paper1}, we have
\begin{eqnarray}
{F_2(x,Q^2) \over x} && = q(x,Q^2)  = { 1 \over 4 \pi} \int d \xi^- 
e^{ - { i \over 2} P^+ \xi^- x}  \sum_{n_1,n_2}
\langle PS \mu^2 \mid n_1 \rangle
\langle n_2 \mid PS \mu^2 \rangle \nonumber \\
&& ~~~~\langle n_1 \mid U_h^{-1}
\Big [ (\psi^+)^\dagger(\xi^-) \psi^+(0) - 
(\psi^+)^\dagger(0) \psi^+
(\xi^-) \Big ] 
U_h \mid n_2 \rangle \, , \label{exp2}
\end{eqnarray}
where $U_h=T^+ \exp\Big\{ -{i\over 2}\int_{-\infty}^0 dx^+ 
\tilde{P}^-_{int}(x^+)\Big\}$, and $\tilde{P}_{int}^- \equiv P_{int}^{-H} 
+ P_{int}^{-M}$ is denoted as the hard and mixed light-front 
interaction Hamiltonian. 

In the following, we shall not explicitly evaluate contributions from
intermediate states which involve vanishing energy denominators. These
contributions are most conveniently included by introducing the wave
function renormalization constant associated with the parton active in 
the high energy process.
 
Let us consider first few terms in the expression given in Eq. (\ref{exp2}).
The lowest order term yields the function 
\begin{eqnarray}
	q^{(0)}(x,Q^2) &=& q(x,\mu^2) = \sum_n |\langle PS, \mu^2_{\rm fact}
		| n\rangle |^2 \nonumber \\
	&=& \sum_s \int^\mu d^2 k^\perp \langle PS \mu^2 \mid 
		b_s^\dagger(y P^+, k^\perp) b_s(yP^+, k^\perp) \mid  
		PS \mu^2 \rangle \, .
\end{eqnarray}
Terms linear in the interaction Hamiltonian vanishes since the 
plus component of the bilocal operator
conserves particle number on the light-front. Consider the second order
contribution, 
\begin{eqnarray}
q^{(2)}(x,Q^2) && =  { 1 \over 4} \sum_{nmpk} \langle PS \mu^2
\mid n \rangle \int _{- \infty}^0 dx_1^+ \langle n \mid
{\tilde P}^{-}_{int}(x_1^+) 
\mid m \rangle  \langle m \mid {\cal O}  
  \mid p \rangle \nonumber \\
&& ~~~~~~~~ \langle p \mid \int _{- \infty}^0 dx_2^+
{\tilde P}^{-}_{int}(x_1^+) \mid k \rangle \langle k \mid P S \mu^2 \rangle.  
\end{eqnarray}
Here we have denoted 
\begin{eqnarray}
{\cal O} = { 1 \over 4 \pi}
\int d \xi^- e^{ - { i \over 2} P^+ \xi^- x}
\Big [ (\psi^+)^\dagger(\xi^-) \psi^+(0) - 
(\psi^+)^\dagger(0) \psi^+
(\xi^-) \Big ]
\end{eqnarray}
Using,
\begin{eqnarray}
P^-_{int}(x^+) = e^{{i \over 2} P_{free}^- (x^+)} P^-_{int}(0)
e^{-{i \over 2} P_{free}^- (x^+)},
\end{eqnarray}
we have,
\begin{eqnarray}
q^{(2)}(x,Q^2) && =  \sum_{nmpk}  
{ 1 \over P^-_{0n} - P^-_{0m} }{ 1 \over P^-_{0k} - P^-_{0p} } 
\langle n \mid {\tilde P}^{-}_{int} (0) \mid m \rangle 
\nonumber \\
&& ~~~~~~~ \langle m \mid {\cal O} \mid p \rangle \langle p \mid 
{\tilde P}^{-}_{int}(0) \mid k
\rangle 
\langle PS \mu^2
\mid n \rangle
\langle k \mid P S \mu^2 \rangle. 
\end{eqnarray}

The states $ \mid n \rangle$, and $ \mid k \rangle$ are forced to be low
energy states with $ (k^\perp)^2  < \mu^2$.  We can restrict the states $
\mid m \rangle$, $ \mid p \rangle$ to be high energy states with $
(k^\perp)^2 > \mu^2$. The bilocal operator $ {\cal O}$ picks an active
parton in a high energy state whose longitudinal momentum is forced to be $
x P^+$. Further we need to keep only terms in ${\tilde P}^{-}_{int} $ 
which cause
transitions involving the active parton. (Transitions involving spectators
lead to wave function renormalization of spectator states which are 
canceled by the renormalization process as shown explicitly
in Sec. VIB.)

To order $\alpha_s$, a straightforward evaluation (see later) leads to 
\begin{eqnarray}
q(x,Q^2) = {\cal N}~ \Big \{ q(x, \mu^2) + { \alpha_s 
\over 2 \pi} C_f ln{Q^2 \over \mu^2}
\int_x^1 { dy \over y} P(x/y) q(y, \mu^2) \Big \}
\end{eqnarray}
where ${\cal N}$ is the wave function renormalization constant 
of the active parton and $P$ is the splitting function. 
Including the contribution from the wave function renormalization
constant to the same order ($\alpha_s$), we get,
\begin{eqnarray}
q(x,Q^2) = \int dy ~{\cal P} (x,Q^2; y, \mu^2) ~ q(y, \mu^2),
\end{eqnarray}
where the hard scattering coefficient 
\begin{eqnarray}
{\cal P} (x,Q^2; y, \mu^2)  =  \delta (x-y) + {
\alpha_s \over 2 \pi} C_f ln{Q^2 \over \mu^2} \int_0^1 dz \delta(zy-x)
{\tilde P}(z)
\end{eqnarray}
with $ {\tilde P}(x) = P(x) - \delta(1-x) \int_0^1 dy P(y)$.

We note that the above analysis can be carried over to all orders in
perturbation theory with the result
\begin{eqnarray}
{\cal P}(x,Q^2; y, \mu^2) &&= \langle yP^+, k^\perp, s \mid U_h^{-1} 
	{\cal O} U_h \mid yP^+, k^\perp, s \rangle \nonumber \\
&& = \langle yP^+, k^\perp, s; (dressed) \mid {\cal O}
      \mid yP^+, k^\perp, s; (dressed)\rangle, \label{hsc}
\end{eqnarray}

In evaluating the above expression, 
only in the interaction Hamiltonians in the extreme
left and extreme right of the time ordered product we need to keep mixture
of soft and hard partons. This is governed by $P^{-M}_{int}$. They are  
needed to cause of transition of a soft parton to a hard parton. In the 
rest of the interaction Hamiltonians occurring in the chain, the partons 
are restricted to be hard, i.e., they are determined by $P^{-H}_{int}$ only. 
For the leading logarithmic evolution we are discussing here, they appear 
ordered in transverse momentum.

\section{Tools of calculation}
In this section, we outline the basic tools for calculating the 
perturbative contribution to the structure functions, namely the
hard scattering coefficients, ${\cal P}(x,Q^2; y, \mu^2)$,
given by Eq.~(\ref{hsc}).  If we set $k=P$, then $y=1$ and the hard
scattering coefficients just become the structure functions of dressed quark
and gluon targets in DIS, 
\begin{equation}
        f_i^p(x,Q^2) = { 1 \over 4 \pi}\int d\eta
                e^{-i\eta x} ~_p\langle ks | \Big[ \overline{\psi} 
		(\xi^-) \Gamma_i \psi (0) \mp h.c. \Big]
                | ks \rangle_p  \, . \label{sqm1}
\end{equation}
As a matter of fact, we can compute the perturbative QCD
correction to the hadronic structure functions by calculating 
the structure functions of the dressed quarks and gluons.
In old-fashioned light-front perturbation theory, the dressed quark
or gluon states can be expanded as follows\cite{Zhang93}: 
\begin{eqnarray}
	\mid Ps \rangle_p && = U_h | Ps\rangle = \sqrt{\cal N} \Big 
		\{ \mid Ps \rangle + \sum'_{n} \mid n \rangle { 
		\langle n \mid P^{-M}_{int} \mid 
		Ps \rangle  \over (P^- - P_n^-) } \nonumber \\
		&& ~~~~~ +  \sum'_{mn} \mid m \rangle{ \langle m \mid 
	P^{-H}_{int} \mid n \rangle \langle n \mid P^{-M}_{int} \mid Ps
	\rangle \over(P^- - P_m^- ) (P^- - P_n^-)} + ... \Big \} \label{exp1}
\end{eqnarray}
where $|Ps \rangle$, the bare single particle state, and $ \mid n 
\rangle$, the two-particle state, $\mid m \rangle$, the three-particle 
state, etc., are eigenstates of the free Hamiltonian.  Single particle 
state $|Ps \rangle$ is omitted in the sum over the states in the 
above expansion. Introducing the multi-parton amplitudes (wave functions),
\begin{eqnarray}
	\Phi_n && = { \langle n \mid H^{-M}_{int} \mid Ps \rangle \over 
		(P^- - P_n^- )}, \nonumber \\
	\Phi_m && = \sum_n{ \langle m \mid P^{-H}_{int} \mid n \rangle 
             \langle n \mid P^{-M}_{int} \mid Ps \rangle \over (P^- - 
		P_m^- ) (P^- - P_n^-)}, 
\end{eqnarray}
the expansion in Eq. (\ref{exp1}) takes the form
\begin{eqnarray}
	\mid P s \rangle_p = \sqrt{{\cal N}} \Big \{ \mid Ps \rangle + 
		\sum_n \Phi_n \mid n \rangle + \sum_m \Phi_m \mid m 
		\rangle + ... \Big \} \, .
\end{eqnarray} 

In the above expressions, $P^{-M}_{int}$ and $P^{-H}_{int}$ are the 
interaction parts of the canonical light-front QCD Hamiltonian, but
the former contains the mixed soft and hard partons and the latter
only has hard partons.  The canonical light-front QCD Hamiltonian is
given by the following form in our two-component formalism \cite{Zhang93}: 
\begin{equation}
	P^-_{int}=\int dx^-d^2x^\bot \Big\{ {\cal H}_{qqg} + {\cal 
		H}_{ggg} + {\cal H}_{qqgg} + {\cal H}_{qqqq} + 
		{\cal H}_{gggg} \Big\} \, ,
\end{equation}
and
\begin{eqnarray}
  {\cal H}_{qqg} = & & g\xi^{\dagger}\biggl \{ - 2\biggl (
   \frac{1}{\,\partial^+\,}\biggr ) (\partial \cdot
  A^{\bot})+\sigma \cdot A^{\bot}\biggl (
   \frac{1}{\,\partial^+\,}\biggr ) (\sigma \cdot \partial^{\bot}
  +m)  \nonumber \\
  & & ~~~~~~~~ + \biggl (  \frac{1}{\,\partial^+\,}\biggr )
  (\sigma \cdot \partial^{\bot}-m)\sigma \cdot A^{\bot}
  \biggr \} \xi \, , \\
  {\cal H}_{ggg} = && g f^{abc} \biggl \{ \partial^i A_a^j A_b^i A_c^j
  + (\partial^i A_a^i) \biggl (  \frac{1}{\,\partial^+\,} \biggr )
  (A_b^j \partial^+ A_c^j) \biggr \} \, , \\
  {\cal H}_{qqgg} = & & g^2 \biggl \{ \xi^{\dagger} \sigma \cdot
  A^{\bot} \biggl (  \frac{1}{\,i\partial^+\,}\biggr ) \sigma \cdot
  A^{\bot}\xi \nonumber \\
  & & ~~~~~~~~ +2\biggl (  \frac{1}{\,\partial^+\,}
  \biggr ) (f^{abc}A_b^i\partial^+A_c^i)\biggl (  \frac{1}{\,\partial^+\,}
  \biggr ) (\xi^{\dagger}T^a\xi ) \biggr \} \nonumber \\
  = && {\cal H}_{qqgg1} + {\cal H}_{qqgg2} \, , \\
  {\cal H}_{qqqq} = && 2g^2 \biggl \{ \biggl (  \frac{1}{\,\partial^+\,}
  \biggr )(\xi^{\dagger}T^a\xi ) \biggl (  \frac{1}{\,\partial^+\,}
  \biggr ) (\xi^{\dagger}T^a\xi ) \biggr \} \, , \\
  {\cal H}_{gggg} = &&  \frac{\,g^2\,}{4}f^{abc}f^{ade}
  \biggl \{ A_b^iA_c^jA_d^iA_e^j+2\biggl (  \frac{1}{\,\partial^+\,}
  \biggr )(A_b^i\partial^+A_c^i)\biggl (  \frac{1}{\,\partial^+\,}
  \biggr ) (A_d^j\partial^+A_e^j)\biggr \} \nonumber \\
  = &&  {\cal H}_{gggg1} + {\cal H}_{gggg2} \, ,
\end{eqnarray}
where the dynamic fermion field operator
\begin{equation}
	\psi^+(x)= \left[\begin{array}{cc} \xi(x) \\ 0 \end{array}
		\right]~~ ,
\end{equation}
with 
\begin{eqnarray}
	\xi(x) &=& \sum_\lambda \chi_\lambda \int {dk^+d^2k^\bot
		\over 2(2\pi)^3 \sqrt{k^+}}\Big(b_\lambda(k)e^{-ikx} + 
		d_{-\lambda}^\dagger(k)e^{ikx} \Big) \, ,
\end{eqnarray}
and the transverse gluon field operator
\begin{eqnarray}
	A^{i \bot}(x) &=& \sum_\lambda \int {dk^+d^2k^\bot\over 
		2(2\pi)^3k^+}\Big(\varepsilon^{i \bot} (\lambda) 
		a_\lambda(k)e^{-ikx} + h.c \Big)
\end{eqnarray}
with
\begin{eqnarray} 
	\Big\{b_\lambda (k), b_{\lambda'}^\dagger(k') \Big\} &=& 
		\Big\{d_\lambda(k), d_{\lambda'}^\dagger{k'} \Big\} 
	 = 2(2\pi)^3 k^+\delta (k^+-{k'}^+) \delta^2(k^\bot - {k^{\bot}}'), \\
	\Big[a_\lambda(k) , a_{\lambda'}^\dagger (k') \Big] &=& 
	 2(2\pi)^3 k^+\delta (k^+-{k'}^+) \delta^2(k^\bot - {k^{\bot}}'), 
\end{eqnarray}
and $\chi_\lambda$ is the eigenstate of $\sigma_z$ in the two-component
spinor of $\psi_+$ by the use of the following light-front $\gamma$ 
matrix representation \cite{Zhang95},
\begin{equation}
	\gamma^0 = \left[\begin{array}{cc} 0 & - i \\ i & 0 \end{array}
		\right] ~~, ~~ 
	\gamma^3=  \left[\begin{array}{cc} 0 & i \\ i & 0 \end{array}
		\right] ~~, ~~
 	\gamma^i =  \left[\begin{array}{cc} -i\tilde{\sigma}^i & 0 \\ 
		0 & i\tilde{\sigma}^i \end{array} \right] 
\end{equation}
with $\tilde{\sigma}^1 =\sigma^2, \tilde{\sigma}^2=-\sigma^1$) and 
$\varepsilon^i(\lambda)$ the polarization vector of transverse 
gauge field.
   
\section{Unpolarized dressed parton structure functions}
In this section we present the calculations of the $F_2$ structure function
for dressed quark and gluon targets.
\subsection{Dressed quark structure function from the plus component}
The $F_2$ structure function of a dressed quark is given by
\begin{eqnarray}
	{F_2(x,Q^2)\over x} &=& {1\over 4\pi} \int d\eta e^{-i\eta x} 
		 \overline{V}_{1} \nonumber  
		 \\
	&=& {1\over 4\pi P^+} \int d\eta e^{-i\eta x}  
		 {}_{p}\langle ks| \overline{\psi} (\xi^-) 
		\gamma^+ \psi (0) - \overline{\psi} (0)
		\gamma^+ \psi (\xi^-) |ks \rangle_p.  \label{f2+}
\end{eqnarray}
The gluon structure function \cite{pQCD} is defined by 
\begin{eqnarray}
	F_2^G(x,Q^2) = { 1 \over 4 \pi P^+} \int d\eta e^{- {i \eta x }} 
		{}_p\langle k \lambda \mid (-)F^{+ \nu a} (\xi^-)F^{+a}_{~~\nu}(0) \mid 
		k \lambda \rangle_p \, .
\end{eqnarray}

The dressed quark or gluon state can be obtained by the perturbative
expansion in the old fashioned time-ordered Hamiltonian formulation,
as given by Eq.~(\ref{exp1}). But we can also find such states by
solving the light-front bound state equation.  Let us take the state 
$\mid P \rangle$ to be a dressed quark which obeys the eigen value 
equation
\begin{equation}
        \Big(M^2 - \sum_{i=1}^n { (\kappa_i^{\bot})^2 + m_i^2 \over x_i} 
		\Big) \left[\begin{array}{c} \Phi_{q} \\
                \Phi_{qg} \\ \vdots \end{array} \right]
                  = \left[ \begin{array}{ccc} \langle q
                | H^H_{int} | q \rangle & \langle q | H^H_{int}
                | qg \rangle & \cdots \\ \langle q g
                | H^H_{int} | q \rangle & \cdots & ~~  \\ \vdots &
                \ddots & ~~ \end{array} \right] \left[\begin{array}{c}
                \Phi_{q} \\ \Phi_{qg} \\ \vdots \end{array}
                \right] . \label{lfbe}
\end{equation}
Explicitly, expanding the state in terms of bare states of quark, quark 
plus gluon, quark plus two gluons, etc, we have,
\begin{eqnarray}
	\mid P \sigma \rangle_q = && \sqrt{{\cal N}_q} \Bigg \{ 
		b^\dagger(P,\sigma) \mid 0 \rangle \nonumber \\
	&& ~ + \sum_{\sigma_1,\lambda_2} \int {dk_1^+ d^2 k_1^\perp \over 
		\sqrt{2 (2 \pi)^3 k_1^+}} \int {dk_2^+ d^2 k_2^\perp 
		\over \sqrt{2 (2 \pi)^3 k_2^+}} \psi_2(P,\sigma  
		\mid k_1, \sigma_1; k_2 , \lambda_2)   \nonumber \\
	&& ~~~~~~~~~~~~~~~~~~~ \times \sqrt{2 (2 \pi)^3 P^+} 
		\delta^3(P-k_1-k_2) b^\dagger(k_1,\sigma_1)
		a^\dagger(k_2,\lambda_2) \mid 0 \rangle \Bigg \}. \label{dsqs}
\end{eqnarray}
The factor  ${\cal N}_q$ is the wave function renormalization constant 
for the quark and the function $\psi_2(P,\sigma \mid k_1 \sigma_1, k_2 
\lambda_2)$ is the probability amplitude to find a bare quark with 
momentum $k_1$ and helicity $\sigma_1$ and a bare gluon 
with momentum $k_2$ and helicity $\lambda_2$ in the dressed quark.

Introduce the Jacobi momenta ($x_i, \kappa_i^\perp$) 
\begin{eqnarray}
k_i^+ = x_i P^+, \,  k_i^\perp = \kappa_i^\perp + x_i P^\perp 
\end{eqnarray} 
so that 
\begin{eqnarray} 
\sum_i x_i = 1, \sum_i \kappa_i^\perp =0.
\end{eqnarray}
The amplitude $ \psi_2$ is related to the amplitude $ \Phi_2$ in Eq.
(\ref{lfbe}) by ,
\begin{eqnarray}
\sqrt{P^+} \psi_2(k_i^+, k_i^\perp) = &&  \Phi_2(x_i,
\kappa_i^\perp).
\end{eqnarray}
The two particle amplitude $ \psi_2$ is given by 
\begin{eqnarray}
	&& \psi_2(P, \sigma \mid p_1,s_1; p_2,\rho_2) = { 1 \over \Big[ {m^2 + 
		(P^\perp)^2 \over P^+} - {m^2 + (p_1^\perp)^2 \over p_1^+} 
		- {(p_2^\perp)^2 \over p_2^+} \Big]} \nonumber \\
	&&~~~~~~~~ \times { g \over \sqrt{2 (2 \pi)^3}} T^a {1 \over 
		\sqrt{p_2^+}}\chi^\dagger_{s_1} \Big[ 2 {p_2^\perp \over 
		p_2^+} - {\sigma^\perp.p_1^\perp - im \over p_1^+}
		\sigma^\perp - \sigma^\perp {\sigma^\perp.P^\perp + im 
		\over P^+}  \Big] \chi_\sigma . {(\epsilon^\perp_{\rho_2})}^*. 
\end{eqnarray}
We rewrite the above equation in terms of Jacobi momenta
($ p_i^+ = x_i P^+, \kappa^\perp_i =p^\perp_i + x_i P^\perp$) and the
wave functions $\Phi_i$ which are functions of Jacobi momenta. Using the 
notation $x=x_1, \kappa_1 = \kappa$ and using the facts $x_1+x_2=1, 
\kappa_1+\kappa_2=0$, we have
\begin{eqnarray}
	&& \Phi_2^{s_1, \rho_2}(x,\kappa^\perp; 1-x, - \kappa^\perp) 
		= { 1 \over \Big[ m^2 - {m^2 +(\kappa^\perp)^2 \over x } -
		{(\kappa^\perp)^2 \over 1-x} \Big] } \nonumber \\
	&&~~~~~~~~~~\times   { g \over \sqrt{2 (2 \pi)^3}} T^a 
		\chi^\dagger_{s_1} \Big[ - 2 {\kappa^\perp \over 1-x} - 
		{\sigma^\perp.\kappa^\perp  -i m \over x} \sigma^\perp - 
		\sigma^\perp i m \Big] \chi_{\sigma} .{(\epsilon^\perp_{
		\rho_2})}^*. \label{qap} 
\end{eqnarray} 

Evaluating the expression in Eq. (\ref{f2+}) explicitly, noting that in the 
present case
the contribution from the second term in this expression is zero,
we get the quark structure function of the dressed quark
\begin{eqnarray}
	{F_{2(q)}^{q}(x,Q^2) \over x} &=& {\cal N}_q \Big \{ \delta (1-x) 
		\nonumber \\
	&& + \sum_{\sigma_1, \lambda_2} \int dx_2 \int d^2 \kappa_1^\perp
	\int d^2 \kappa_2^\perp \delta (1-x-x_2) \nonumber \\
	&& ~~~~~~~\times \delta^2(\kappa_1^\perp+\kappa_2^\perp) 
		\mid \Phi_2^{\sigma_1, \lambda_2}(x,\kappa_1^\perp; x_2,
	\kappa_2^\perp) \mid^2 \Big \}. \label{f2plus}
\end{eqnarray}
This equation makes manifest the parton interpretation of the
quark distribution function, namely, the quark distribution function of a
dressed quark is the
incoherent sum of probabilities to find a bare parton (quark) with longitudinal
momentum fraction $x$ in various multi-particle Fock states of the dressed 
quark. Since we have computed the distribution function in field theory,
there are also significant
differences from the traditional parton model\cite{Feynman72}. Most important
difference is the fact that the partons in field theory have transverse
momenta ranging from zero to infinity. Whether the structure function scales
or not now depends on the ultraviolet behavior of the multi-parton
wave functions. By analyzing various interactions, one easily finds that in
super renormalizable interactions, the transverse momentum integrals 
converge in the ultraviolet and the structure function scales, 
whereas in renormalizable
interactions, the transverse momentum integrals diverge in the ultraviolet
which in turn leads to scaling violations in the structure function. 

Taking the bare and dressed quarks to be massless,
we arrive at
\begin{eqnarray} 
	&& \sum_{\sigma_1,\rho_2} \int d^2 \kappa^\perp \mid
		\Phi_2^{\sigma_1,\rho_2}(x,\kappa^\perp, 1-x, 
		-\kappa^\perp) \mid^2 \nonumber \\
	&&~~~~~~~~~~~~~~~= {g^2 \over (2 \pi)^3}  C_f { 1 + x^2 \over 1-x} 
	\int d^2 \kappa^\perp { 1 \over (\kappa^\perp)^2} \label{qdens}
\end{eqnarray}    
where $C_f = {N^2 -1 \over 2N}$. Recalling that 
$\mid \Phi_2(x, \kappa^\perp) \mid^2$ is the probability density to
find a quark with momentum fraction $x$ and relative transverse momentum
$\kappa^\perp$ in a parent quark, we define the probability density to find
a quark with momentum fraction $x$ inside a parent quark as
the splitting function 
\begin{eqnarray}
	P_{qq}(x)= C_f {1 + x^2 \over 1-x}.
\end{eqnarray}
Clearly, the probability density to find a gluon with momentum fraction 
$x$ inside a parent quark is defined as the splitting function 
\begin{eqnarray}
	P_{Gq}(x) = C_f {1 + (1-x)^2 \over x}.  
\end{eqnarray}

The transverse momentum integral in Eq. (\ref{qdens}) is divergent at 
both limits of integration.  We regulate the lower limit by $\mu$ and the 
upper limit by $Q$. Thus we have 
\begin{eqnarray}
	{F_{2(q)}^q(x,Q^2) \over x} = {\cal N}_q  \Big [ \delta (1-x) \, 
		+ \, {\alpha_s \over 2 \pi}C_f{1+x^2 \over 1-x} \ln{Q^2 
		\over \mu^2} \Big].
\end{eqnarray}
The normalization condition reads 
\begin{eqnarray}
	{\cal N}_q \Big[ 1 + {\alpha_s \over 2 \pi} C_f
		\int dx{ 1 + x^2 \over 1-x}
		\ln{Q^2 \over \mu^2} \Big ] = 1.
\end{eqnarray}
Within the present approximation (valid only up to $\alpha_s$),
\begin{eqnarray}
	{\cal N}_q = 1 - {\alpha_s \over 2 \pi} C_f \int dx{ 1 + 
		x^2 \over 1-x} \ln{Q^2 \over \mu^2} .
\end{eqnarray}
In the second term we recognize the familiar expression of wave function 
correction of the state $n$ in old fashioned perturbation theory, 
namely,  $ \sum'_m {\mid \langle m \mid V \mid n \rangle \mid^2 \over 
(E_n -E_m)^2}$.

Thus to order $\alpha_s$,
\begin{eqnarray}
	{F_{2(q)}^{q}(x,Q^2) \over x} = \delta(1-x) + {\alpha_s \over 2 \pi}  
		ln{Q^2 \over \mu^2} ~C_f~\Big[ {1+x^2 \over 1-x} - 
		\delta(1-x) \int dy {1+y^2 \over 1-y} \Big]. \label{onel}
\end{eqnarray}
Note that  (\ref{onel}) can also be written as
\begin{eqnarray}
	{F_{2(q)}^q(x,Q^2) \over x} = \delta(1-x) + {\alpha_s \over 2 \pi} 
		C_f ln{Q^2 \over \mu^2}\Big[ {1 +x^2 \over (1-x)_{+}} + 
		{3 \over 2} \delta(1-x) \Big], \label{f2qq}
\end{eqnarray}
which is a more familiar expression. By construction, $\mid \Phi_2(x, 
\kappa^\perp) \mid^2$ is a probability density. However, this function is 
singular as $x \rightarrow 1$ (gluon longitudinal momentum fraction 
approaching zero). To get a finite probability density we have to 
introduce a cutoff $\epsilon$  ($x_{gluon} > \epsilon)$, for example. In 
a physical cross section, this $\epsilon $ cannot appear and here we have 
an explicit example of this cancellation. 
Note that the function 
${\tilde P}_{qq} =  C_f{1 +x^2 \over (1-x)_{+}} + {3 \over 2} \delta(1-x) $
does not have the probabilistic interpretation since it includes contribution
from virtual gluon emission. This is immediately transparent from the
relation
\begin{eqnarray}
	\int dx {\tilde P}_{qq}(x) = 0.
\end{eqnarray}
 We also note that the divergence arising from small transverse momentum
(the familiar mass singularity) cannot be handled properly in the present 
calculation. This is to be contrasted
with the calculation of the physical hadron structure function 
where the mass singularities can be properly absorbed into the
non-perturbative part of the structure function.

Let us now explicitly check the longitudinal momentum sum rule for the
dressed quark. According to the sum rule, 
\begin{eqnarray}
	\int_0^1 dx \Big [ F_{2(q)}^q(x) + F_{2(G)}^q(x) \Big ] = 
		{ 1 \over 2 (P^+)^2} ~{}_q\langle
		P \mid \theta^{++}(0) \mid P \rangle_q =1.
\end{eqnarray}
Explicit calculations show that the gluon structure function for the dressed
quark target $F_{2(G)}^{q}$ is given by
\begin{eqnarray}
	F_{2(G)}^{q} = { \alpha_s \over 2 \pi} ln{Q^2 \over \mu^2} C_f x {1 +
		(1-x)^2 \over x} \label{f2qg}    
\end{eqnarray}
From Eqs. (\ref{f2qq}) and (\ref{f2qg}) it follows that
\begin{eqnarray}
	\int_0^1 dx F_2^q (x) = \int_0^1 dx \Big [ F_{2(q)}^q(x) + 
		F_{2(G)}^q(x) \Big ] = 1
\end{eqnarray}
since
\begin{eqnarray}
	\int_0^1 dx ~x \Big [ {\tilde P}_{qq}(x) + P_{Gq}(x) \Big ] =0.
\end{eqnarray}

\subsection{Dressed  quark structure function from the
transverse component}
From BJL expansion and light-front current algebra, it also follows that
\begin{eqnarray}
	{F_2(x,Q^2)\over x} &=& {1\over 4\pi} \int d\eta e^{-i\eta x} 
		 \overline{V}_{1} \nonumber 
		 \\
	&=& {1\over 4\pi P^{i\bot}} \int d\eta e^{-i\eta x} 
                 {}_p\langle ks | \overline{\psi} (\xi^-) 
		\gamma^i \psi (0) - \overline{\psi} (0)
		\gamma^i \psi(\xi^-) |ks \rangle_p \, . \label{f2i}
\end{eqnarray}

From Eqs.~(\ref{f2+}) and (\ref{f2i}), it follows that the 
structure function $F_2$ can be expressed not only as a matrix 
element of the plus component of the bilocal vector current, but also
the matrix element of the transverse component of the bilocal
current. Next we extract the structure function $F_2(x)$ from 
the transverse component of the bilocal vector current (Eq.~(\ref{f2i})). 
The operator that appears in this equation is
\begin{eqnarray}
{\overline \psi}(y) \gamma^\perp \psi(0) = (\psi^+)^\dagger (y) \alpha^\perp
\psi^-(0) + (\psi^-)^\dagger(y)  \alpha^\perp \psi^+(0). \label{perp}
\end{eqnarray}
The constrained fermion field 
$ \psi^- = {1 \over i \partial^+} (\alpha^\perp.(i \partial^\perp + g
A^\perp) + \gamma^0 m) \psi^+$.
Hence the operator in the above equation appears to be higher twist (twist
three).
Without loss of generality we take the $\perp$ direction along the $x$ axis.
The structure function can be explicitly written as
\begin{eqnarray}
	{F_2(x,Q^2) \over x} &=& { 1 \over 8 \pi} {P^+ \over P^1} \int dy^- 
	e^{- {i \over 2} P^+ y^- x} \langle P \mid \xi^\dagger(y) 
		\nonumber \\
	& & ~~~~~~~~ \times \Big [O_m + O_{k^{\perp}} + O_g \Big ] 
		\xi(0) \mid P \rangle + h.c   , 
\end{eqnarray}
with
\begin{eqnarray}
	O_m && = im { 1 \over i \partial^+} \sigma^2, \nonumber \\
	O_{k^{\perp}} && = { 1 \over i \partial^+} \big [ i \partial^1 
		- \sigma^3 \partial^2 \big], \nonumber \\
	O_g && = g { 1 \over i \partial^+} \big [ A^1 + i \sigma^3 A^2 ] \big .
\end{eqnarray}

First consider contribution from the operator $O_m$. Only potential 
non-vanishing contributions are from
the diagonal matrix elements for the single quark state and the quark-gluon
state. Single quark matrix element vanishes because
$\sigma^2$ flips helicity. Diagonal contribution from the quark-gluon state
also vanishes because of the cancellation between the two terms in
Eq. (\ref{perp}).
Thus the contribution from $O_m$ to $F_2$ vanishes.

Next consider contribution from the operator $O_{k^\perp}$.
 Explicit evaluation leads to  
\begin{eqnarray}
	{F_2(x,Q^2) \over x}\mid_{k^\perp} &=& {\cal N}_q \Big \{ \delta
		(1-x) \nonumber \\
	&& + {1 \over P^1}  \sum_{\sigma_1,\lambda_2} \int dx_2 \int d^2 
		\kappa_1^\perp \int d^2 \kappa_2^\perp \delta(1-x-x_2) 
		\delta^2(\kappa_1^\perp + \kappa_2^\perp) \nonumber \\
	&& ~~~~~~~~~~~ \times \mid \Phi_2^{\sigma_1, \lambda_2} 
		(x,\kappa_1^\perp; x_2, \kappa_2^\perp)\mid^2  { \kappa_1^1 
		+ x P^1 \over x} \Big \} \nonumber \\
	&=& {\cal N}_q \Big \{ \delta (1-x) \nonumber \\
	&& +  \sum_{\sigma_1,\lambda_2} \int dx_2 \int d^2 \kappa_1^\perp 
		\int d^2 \kappa_2^\perp \delta(1-x-x_2) \nonumber \\
	&& ~~~~~~\times \delta^2(\kappa_1^\perp + \kappa_2^\perp) 
		\mid \Phi_2^{\sigma_1, \lambda_2} (x,\kappa_1^\perp; x_2, 
		\kappa_2^\perp) \mid^2 \Big \} \label{kperp} 
\end{eqnarray}
since $
\int d^2 \kappa_1^\perp \kappa_1^1 \mid \psi_2 \mid^2 =0$
as a consequence of rotational invariance. Eq.(\ref{kperp}) gives
the same result as Eq. (\ref{f2plus}).

Lastly we evaluate the contribution from the quark-gluon correlation
operator $O_g$.
\begin{eqnarray}
	{F_2(x,Q^2) \over x} \mid_g && = { 1 \over 2} 
		{ g \over \sqrt{2 (2 \pi)^3 }} { 1 \over P^1} 
		\sum_{\sigma_1, \lambda_2} \int {dy \over \sqrt{1-y}} d^2
		\kappa^\perp \chi^\dagger_\sigma \big [ \epsilon^1_{\lambda_2}
 		+ i \sigma^3 \epsilon^2_{\lambda_2} \big ] \chi_{\sigma_1} 
		\nonumber \\
		&& ~~~~~~~~~~~~\Phi_2^{\sigma_1,\lambda_2}(y, \kappa^\perp; 
			1-y, -\kappa^\perp) + h.c. \nonumber \\
		&& = 0.
\end{eqnarray}
This is because the quark-gluon amplitude $\Phi_2$ has two types 
of  terms: a) terms proportional to the quark mass $m$ accompanied 
by $\sigma^\perp$ which vanish because $ \chi^\dagger_\sigma
\sigma^\perp \chi_\sigma = 0$, and  b) 
terms proportional to $\kappa^\perp$ which vanish because of rotational
symmetry. Thus the contribution from $O_g$ to the structure function
vanishes.

From Eq. (\ref{kperp}) and Eq. (\ref{f2plus}), it follows that 
the structure function extracted from Eq. (\ref{f2i}) has the 
same result given by Eq. (\ref{onel}) and hence the same parton 
interpretation as that extracted from Eq. (\ref{f2+}). 
Thus we have explicitly demonstrated the parton interpretation of the
transverse component of the bilocal vector matrix element in unpolarized
deep inelastic scattering. The classification of twist in 
DIS or other hadronic collision processes based on the different 
components of light-front bilocal operators seems unreliable. 

\subsection{Dressed gluon structure function}
The dressed gluon state can be expanded as
\begin{eqnarray}
	\mid P \lambda \rangle_g =&& {\sqrt{\cal N}_g} \Big \{
		a^{\dagger}(P, \lambda) \mid 0 \rangle  \nonumber \\
	&& + \sum_{\sigma_1 \sigma_2} \int 
		{ dk_1^+ d^2 k_1^\perp \over \sqrt{2 (2 \pi)^3 k_1^+}}
		{ dk_2^+ d^2 k_2^\perp \over \sqrt{2 (2 \pi)^3 k_2^+}}
		\sqrt{2 (2 \pi)^3 P^+} \delta^3(P-k_1-k_2) \nonumber \\
	&& ~~~~ \psi_{2(q {\bar q})}(P, \lambda \mid k_1 \sigma_1, k_2 
		\sigma_2) b^\dagger(k_1 \sigma_1) d^\dagger(k_2, \sigma_2)  
		\mid 0 \rangle \nonumber \\
	&& + { 1 \over 2} \sum_{\lambda_1 \lambda_2} \int 
		{ dk_1^+ d^2 k_1^\perp \over \sqrt{2 (2 \pi)^3 k_1^+}}
		{ dk_2^+ d^2 k_2^\perp \over \sqrt{2 (2 \pi)^3 k_2^+}}
		\sqrt{2 (2 \pi)^3 P^+} \delta^3(P-k_1-k_2) \nonumber \\
	&& ~~~~ \psi_{2(gg)}(P, \lambda \mid k_1 \lambda_1, k_2 \lambda_2)
	a^\dagger(k_1 \lambda_1) a^\dagger(k_2, \lambda_2)  \mid 0 \rangle 
		\Big \}. \label{stateg}
\end{eqnarray}
The factor ${ 1 \over 2}$ is the symmetry factor for identical bosons.

As before we introduce the boost invariant amplitudes
\begin{eqnarray}
	\sqrt{P^+}\psi_{2(q {\bar q})}(k_i^+, k_i^\perp) &=&  \Phi_{2(q
		{\bar q})} (x_i,\kappa_i^\perp), \nonumber \\
	\sqrt{P^+}\psi_{2(gg)}(k_i^+, k_i^\perp) &=&  \Phi_{2(gg)} (x_i,
		\kappa_i^\perp).
\end{eqnarray}
The $ q {\bar q}$ wave function of the dressed gluon is given by
\begin{eqnarray}
	& & \Phi_2^{s_1, s_2}(x,\kappa^\perp; 1-x, - \kappa^\perp) 
		= { 1 \over \Big[ m^2 - {m^2 +(\kappa^\perp)^2 \over 
		x (1-x)}  \Big] } \nonumber \\ 
	& & ~~~~~~~\times   { g \over \sqrt{2 (2 \pi)^3}} T^a 
		\chi^\dagger_{s_1} \Big[ {\sigma^\perp.\kappa^\perp 
		\over x} \sigma^\perp - \sigma^\perp
		{\sigma^\perp .\kappa^\perp \over 1-x}- i { m \over x 
		(1-x)} \sigma^\perp \Big] \chi_{-s_{2}} 
		{(\epsilon^\perp_{\rho_2})}^*. \label{gap} 
\end{eqnarray} 

The $gg$ wave function of the dressed gluon state given by
\begin{eqnarray}
	\Phi_{2(gg)}(x, \kappa^\perp) =&& { g \over \sqrt{2 (2 \pi)^3}} 
		2 i f^{abc} { x(1-x) \over (\kappa^\perp)^2} { 1 \over 
		\sqrt{x}} { 1 \over \sqrt{1-x}} \nonumber \\
	&& ~~~ \epsilon^j_{\lambda_1} \epsilon^l_{\lambda_2}
		(\epsilon^i_{\lambda})^*\Big [ - \kappa^i \delta_{lj} + 
		{ \kappa^j \over x} \delta_{il} + {\kappa^l
		\over 1-x} \delta_{ij} \Big ],
\end{eqnarray}

The contribution from the first term in Eq. (\ref{stateg}) to the gluon
structure function for the dressed gluon target is given by
\begin{eqnarray}
	F_{2(G)}^{g(1)} =  \delta(1-x).
\end{eqnarray}
The contribution to the gluon structure function 
from the $q {\bar q}$ component of the dressed gluon state
is a disconnected contribution which we omit.
The contribution to the gluon structure function
from the $gg$ component of the dressed gluon state is given by 

\begin{eqnarray}
	F_{2(G)}^{g(3)}= {\alpha_s \over 2 \pi} \ln{Q^2 \over \mu^2} 2 N 
		\Big [ { x \over 1-x} + { 1-x \over x} + x(1-x) \Big ]x.
\end{eqnarray}
We define the probability density to a find a gluon with momentum fraction
$x$ in the dressed gluon, $P_{GG}(x)$ by 
\begin{eqnarray}
	P_{GG}(x) = 2 N \Big [ {x \over 1-x} + {1-x \over x} + 
		x (1-x) \Big ].
\end{eqnarray}
Collecting the three contributions together, we have,
\begin{eqnarray}
	F_{2(G)}^g(x,Q^2) =  {\cal N}_g \Big [ \delta(1-x) + {\alpha_s 
		\over 2 \pi} \ln{Q^2 \over \mu^2} 2N 	
		[{ x \over 1-x} + { 1-x \over x} + x (1-x)]x  \Big].
\end{eqnarray} 

The coefficient $ {\cal N}_g$ is determined from the longitudinal 
momentum sum rule for the dressed gluon target, namely, we require,
\begin{eqnarray}
	\int_0^1 dx F_2^g(x) = \int_0^1 dx \Big [ F_{2(G)}^g(x) 
		+ F_{2(q)}^g(x) \Big ] ={ 1 \over 2 (P^+)^2} ~{}_g\langle P 
		\mid \theta^{++}(0) \mid P \rangle_g =1 \label{normg}.
\end{eqnarray}
Thus we need to evaluate
\begin{eqnarray}
{ 1 \over 2 (P^+)^2} ~ {}_g\langle P \mid \theta^{++}_q(0) \mid P \rangle_g.
\end{eqnarray}
Explicit evaluation leads to 
\begin{eqnarray}
{ 1 \over 2 (P^+)^2} {}_g\langle P \mid \theta^{++}_q(0) \mid P \rangle_g =
{\alpha_s \over 2 \pi} ln{Q^2 \over \mu^2} { 1 \over 2} \int dx \Big [ x^2 +
(1-x)^2 \Big] {\cal N}_g.
\end{eqnarray}
We define the probability density to find a quark with momentum fraction $x$
in a dressed gluon as the splitting function $P_{qG}(x)$ 
\begin{eqnarray}
P_{qG}(x) = {1 \over 2} \Big [ x + (1-x)^2 \Big ]. 
\end{eqnarray}

From Eq. (\ref{normg}) we arrive at 
\begin{eqnarray}
	{\cal N}_g \Big [ 1 + {\alpha_s \over 2 \pi}\ln{Q^2 \over \mu^2} 
		\int dx \Big \{  [x^2 + (1-x)^2 ]+ 2N [ { x \over 1-x} 
		+ { 1-x \over x} + x (1-x)]x \Big\} \Big] =1.
\end{eqnarray}
Thus to order $ \alpha_s$, we have 
\begin{eqnarray}
	{\cal N}_g =1 - {\alpha_s \over 2 \pi} \ln{Q^2 \over \mu^2} 
		\int dx \Big \{  [x^2 + (1-x)^2 ]+ 2N 
	[ { x \over 1-x} + { 1-x \over x} + x (1-x)]x \Big\}.
\end{eqnarray}
Correspondingly, the complete dressed gluon structure function is
given by 
\begin{eqnarray}
	F_{2(G)}^g(x,Q^2) &&= \delta(1-x) + {\alpha_s \over 2 \pi} 
		\ln{Q^2 \over \mu^2} \nonumber \\
	&&~~~\Big \{ 2N \Big [ [{x \over (1-x)_+} + { 1-x \over x} + 
		x (1-x)]x + { 11 \over 12}\delta(1-x) \Big ] - { 1 \over 3} 
		\delta (1-x) \Big \}. 
\end{eqnarray}
Including the end point ($ x \rightarrow 1 $) contributions, we define,
\begin{eqnarray}
	{\tilde P}_{GG}(x) = 2N \Big \{ \Big [ { x \over (1-x)_+} 
		+ { 1-x \over x} + x (1-x) \Big ] + { 11 \over 12} 
		\delta(1-x) \Big \} - { 1 \over 3} \delta(1-x).
\end{eqnarray}
To the best of our knowledge, this is the first time gluon
splitting funtion has been calculated using multi-parton 
wave-funtions. There exist some discussions in the literature 
regarding the calculation of DIS splitting funtions using the 
language of multi-parton wave-funtions mainly due to Lepage and 
Brodsky \cite{Brodsky}. But for the gluon splitting funtion, 
they have simply quoted the result from Altarelli-Parisi paper 
\cite{AP}. It is easy to verify that
\begin{eqnarray}
	\int_0^1 dx ~x  \Big [ 2P_{qG}(x) + {\tilde P}_{GG}(x) \Big ] =0.
\end{eqnarray}  

\section{Polarized dressed parton structure functions}
In this section we discuss in detail the calculations of polarized 
structure functions of a dressed quark target in 
light-front perturbation theory.
\subsection{Chirality structure function}
The chirality structure function of a dressed quark is given by
\begin{eqnarray}
	g_1(x,Q^2) &=& {1\over 8\pi} \int d\eta e^{-i\eta x} 
		 \Big(\overline{A}_{1} + {1 \over 2}
           P^+ \xi^- \overline{A}_{2} \Big) \label{g10} \\ 
	&=& {1\over 8 \pi S^+} \int d\eta e^{-i\eta x} 
		 {}_p\langle k s| \overline{\psi} (\xi^-)
		\gamma^+ \gamma_5 \psi(0) + \overline{\psi}(0) 
		\gamma^+\gamma_5 \psi(\xi^-) |ks \rangle_p . \label{g1}
\end{eqnarray}
On the light-front, chirality and intrinsic helicity of a fermion coincide.
A direct calculation of $g_1(x)$ from Eq. (\ref{g1}) in the free quark 
helicity state immediately leads to the well-known solution:
\begin{equation}
	g_1(x) =  {e_q^2\over 2} \delta(1-x) .
\end{equation}
By a calculation similar to that of $F_2$, for $g_1$,  we have for the
dressed quark,
\begin{equation}  \label{g1s}
	g_1(x,Q^2) = {e^2_q \over 2} \Bigg\{\delta(1-x) + {\alpha_s 
		\over 2\pi} C_f \ln{Q^2\over \mu^2} \Bigg[{1+x^2 
		\over (1-x)_+} + {3\over 2}\delta(1-x) \Bigg]\Bigg\} ,
\end{equation}
which has the same form as Eq.~(\ref{f2qq}). Therefore, the splitting
function for $g_1$ is the same as that for $F_2$.

\subsection{Transverse polarized structure function}
The transversely polarized
structure function of a dressed quark is given by
\begin{eqnarray}
	g_T(x,Q^2) &=& {1 \over 8\pi} \int d\eta e^{-i\eta x} 
		 \overline{A}_{1} 
		\label{gt0}\\
	&=& {1\over 8\pi S^i_T} \int d\eta 
		e^{-i\eta x}  {}_p\langle ks|\overline{
		\psi}(\xi^-) \Big(\gamma^i -{P^i\over P^+}
		\gamma^+ \Big)\gamma_5 \psi(0) +~ h.c. |ks \rangle_p 
		\, . \label{gt}
\end{eqnarray}

For convenience, we take the polarization along the $x$-direction.
The transverse polarized quark target in the $x$-direction 
can be expressed in terms of helicity states by 
\begin{equation}  \label{thb}
	| k^+, k_\bot, S^1 \rangle = {1\over \sqrt{2}}\Big(| k^+, 
		k_\bot, \uparrow \rangle \pm | k^+, k_\bot, \downarrow 
		\rangle \Big)
\end{equation}
with $S^1=\pm m^R_q$, and $m^R_q$ is the renormalized quark mass.
The operator (see Eq. (\ref{gt})) 
\begin{equation}
	\overline{\psi} (\xi^-) \gamma_\bot 
		  \gamma_5 \psi (0)
		= O^m + O^{k_\bot} + O^g 
\end{equation}
where
\begin{eqnarray}
	&& O^m = m \psi_+^\dagger (\xi^-) \gamma_\bot \Big({1 \over i 
		\roarrow{\partial}^+} - {1\over i \loarrow{\partial}^+} 
		\Big)\gamma_5 \psi_+(0) \, , \nonumber \\
	&& O^{k_\bot}= -\psi_+^\dagger (\xi^-)\Big(\gamma_\bot {1\over
		\roarrow{\partial}^+}{\not \! \roarrow{\partial_\bot}} 
		+ {\not \! \loarrow{\partial_\bot}}{1\over \loarrow{
		\partial}^+}\gamma_\bot \Big) \gamma_5\psi_+(0) \, ,
		\nonumber \\
	&& O^g = g\psi_+^\dagger(\xi^-) \Big ({\not \! \! A_\bot}(\xi^-)
		{1\over i\loarrow{\partial}^+}\gamma_\bot - \gamma_\bot 
		{1\over i\roarrow{\partial}^+}{\not \! \! A_\bot}(0) 
		\Big)\gamma_5 \psi_+(0) ~ \label{go}
\end{eqnarray} 
where $m$ and $g$ are the quark mass and quark-gluon coupling 
constant in QCD, and $A_\bot=A^a_{\bot}T_a$ is the transverse 
gauge field. 

Without the QCD correction (i.e., for the free quark state), it 
is easy to show that  
\begin{equation}
	g_T(x) =g_T^m(x) ={e_q^2\over 2} {m_q \over S^1} \delta(1-x)
		= {e_q^2\over 2} \delta(1-x), 
	~~~ g_T^{k_\bot}(x)=0=g_T^g(x).
\end{equation}
Here $m_q/S^1 = 1$ since the renormalized mass is the same as the 
bare mass at the tree level of QCD. We see that only the quark mass 
term contributes to $g_T$ in Eq. (\ref{go}). The quark transverse
momentum term alone does not contribute to $g_T$ since it cannot
cause helicity flip in the free theory. This result indicates that 
physically the dominant contributions to $g_T$ is not controlled 
by the twist classification. 
Thus, for the free quark, we have 
\begin{equation}  \label{fqg2}
	g_2(x)=g_T(x) - g_1(x) = 0 .
\end{equation} 
It is obvious that for free theory, the Burkhardt-Cottingham
(BC) sum rule is trivially obeyed. But as we can see (as it has also 
been previously noticed) the Wandzura-Wilczek relation \cite{WW},
\begin{equation}
	g_2(x) = - g_1(x) + \int_x^1 dy {g_1(y)\over y},
\end{equation}
is not satisfied in free theory. 

Next, we consider the QCD corrections up to order $\alpha_s$, where
the quark-gluon interaction is explicitly included.  We find that all 
the three terms in Eq. (\ref{go}) have nonzero contribution 
to $g_T$, 
\begin{eqnarray}
	g_T^m(x,Q^2) &=&{e^2_q\over 2} {m_q\over S^1}
 		\Bigg\{\delta(1-x) + {\alpha_s \over 2\pi} 
		C_f \ln{Q^2\over \mu^2} \Bigg[{2 \over 
		1-x} \nonumber \\
	&& ~~~~~~~~~~~~~~  -~\delta(1-x) \int_0^1 dx'
		{1+x'^2\over 1-x'} \Bigg]\Bigg\}, \label{ems}\\
	g_T^{k_\bot}(x,Q^2) &=&-{e^2_q\over 2} {m_q\over S^1} 
		{\alpha_s \over 2\pi} C_f \ln{Q^2\over \mu^2}(1-x) , 
		\label{gtk} \\
	g_T^g(x,Q^2) &=& {e^2_q\over 2} {m_q\over S^1}{\alpha_s 
		\over 2\pi} C_f \ln{Q^2\over \mu^2}{\delta(1-x) \over 2}, 
		\label{gtg}
\end{eqnarray}
from the dressed quark wave function normalization constant ${\cal N}$ 
in Eq. (\ref{dsqs}) (corresponds to the virtual contribution in the 
standard Feynman diagrammatic approach). It shows that up to the 
order $\alpha_s$, the matrix elements from $O_{k_\bot}$ (quark 
transverse momentum effect) and $O_g$ (quark-gluon interaction
effect) in Eq. (\ref{go}) are also proportional to quark mass. 
In other words, the transverse quark momentum and quark-gluon 
coupling contributions to $g_T(x,Q^2)$ arise from quark mass effect. 
Explicitly, these contributions arise from the interference of the 
$m_q$ term with the non-$m_q$ dependent terms in the wave function
of Eq. (\ref{qap}) through the quark transverse momentum operator
and the quark-gluon coupling operator in the $g_T$ expression.
This result is not surprising since, as we have pointed out, the 
pure transverse polarized structure function measures the dynamical 
effect of chiral symmetry breaking \cite{Zhang96}. Physically 
only those interferences related to quark mass can result in
the helicity flip (i.e., chiral symmetry breaking) in pQCD so 
that they can contribute to $g_T(x,Q^2)$. From this result, we 
may see that only the operators themselves or their twist structures 
may not give us useful information about their importance in 
the determination of structure functions.

Combining the results together, we obtain 
\begin{equation}
	g_T(x,Q^2) = {e^2_q \over 2} {m_q \over S^1} \Bigg\{
		\delta(1-x) + {\alpha_s \over 2\pi} C_f \ln{Q^2\over 
		\mu^2} \Bigg[{1+2x-x^2 \over (1-x)_+} + 2 \delta(1-x) 
		\Bigg] \Bigg\} . \label{gt1}
\end{equation}
Note that in the above solution, $m_q$ is the bare quark mass,
while the dressed quark polarization $S^1= m^R_q$, and
up to order $\alpha_s$, 
\begin{equation}
	m_q^R = m_q \Bigg( 1 + {3\over 4\pi} \alpha_s C_f \ln
		{Q^2\over \mu^2} \Bigg) .
\end{equation}
We must emphasize that on the light-front there are two mass
scales in the QCD Hamiltonian, one is proportional to $m_q^2$
which does not violate chiral symmetry, and the other is 
proportional to $m_q$ which we discuss here and is associated
with explicit chiral symmetry breaking in QCD \cite{Wilson94}.
 An important
feature of light-front QCD is that the above two mass scales
are renormalized in different ways even in the perturbative 
region.  The renormalization of $m_q^2$ in pQCD is different
from the above result, the details of which can be found in
 our previous work \cite{Zhang93}. 
With this consideration, we have
\begin{equation}
	g_T(x,Q^2) = {e^2_q\over 2}  \Bigg\{\delta(1-x) + 
		{\alpha_s \over 2\pi} C_f \ln{Q^2\over \mu^2} 
		\Bigg[{1+2x-x^2 \over (1-x)_+} + {1\over 2} 
		\delta(1-x) \Bigg] \Bigg\} . \label{gts}
\end{equation}
The final result is independent quark mass but we have to emphasize
again that one must start with massive quark theory\cite{Alt}. Otherwise, 
there is no definition for $g_T$ at the beginning. This result has a close
analogy in helicity flip process at high energy
in Quantum Electro Dynamics\cite{Sehgal}. Cross section for such process,
which vanish in the chiral limit according to naive arguments, indeed is
non-vanishing if one lets the electron mass to go zero at the end of the
calculation as Lee and Nauenberg pointed out long time ago \cite{Lee}.   
 
 Thus, 
up to of the oder $\alpha_s$, we find $g_2$ for a quark target 
\begin{equation}
	g_2(x,Q^2) = {e_q^2\over 2}{\alpha_s \over 2\pi} C_f 
		\ln{Q^2\over \mu^2} \Big[2x - \delta(1-x) \Big].
\end{equation}
It is easy to check that the above result of $g_2(x,Q^2)$ obeys 
the BC sum rule,
\begin{equation}
	\int_0^1 dx g_2(x,Q^2) = 0,
\end{equation}
as is expected. Our results also show that Wandzura-Wilczek relation is
strongly violated in perturbative QCD.

\section{Structure Function of hadron: Parton picture, Scale evolution and
Factorization}

As we have schematically discussed in Sec.~II (also see \cite{paper1}),
the nonperturbative contribution to the structure functions and the 
scaling violations from the perturbative QCD corrections can be unified and 
treated in the same framework in our formalism. In this section, we shall 
address the issues associated with scaling  violations in the structure 
function of the ``meson-like" bound state and explicitly demonstrate the
validity of factorization outlined in Sec. II.
\subsection{Parton picture}
Let us first discuss the emergence of parton picutre for the structure
function of a composite state.
We expand the state $\mid P \rangle $ for $ q {\bar q}$ bound state in 
terms of the Fock components $q {\bar q}$, $q {\bar q}g$, ... as follows.
\begin{eqnarray}
\mid P \rangle = && \sum_{\sigma_1, \sigma_2} 
\int {dk_1^+ d^2 k_1^\perp \over \sqrt{2 (2 \pi)^3 k_1^+}} 
\int {dk_2^+ d^2 k_2^\perp \over \sqrt{2 (2 \pi)^3 k_2^+}} 
\nonumber \\
&& \psi_2(P \mid k_1, \sigma_1; k_2, \sigma_2) \sqrt{2 ((2 \pi)^3 P^+}
\delta^3(P-k_1-k_2) b^\dagger(k_1, \sigma_1) d^\dagger(k_2,\sigma_2) \mid 0
\rangle \nonumber \\
&& + \sum_{\sigma_1,\sigma_2,\lambda_3} 
\int {dk_1^+ d^2 k_1^\perp \over \sqrt{2 (2 \pi)^3 k_1^+}} 
\int {dk_2^+ d^2 k_2^\perp \over \sqrt{2 (2 \pi)^3 k_2^+}} 
\int {dk_3^+ d^2 k_3^\perp \over \sqrt{2 (2 \pi)^3 k_3^+}} 
\nonumber \\
&& \psi_3(P \mid k_1, \sigma_1; k_2, \sigma_2; k_3, \lambda_3)
\sqrt{2 (2 \pi)^3 P^+} \delta^3(P-k_1 -k_2 -k_3) 
\nonumber \\
&&~~~~~b^\dagger(k_1 ,\sigma_1)
d^\dagger (k_2, \sigma_2) a^\dagger(k_3, \lambda_3) \mid 0 \rangle \nonumber
\\
&& + \, \, \, ... \, \, \, . \label{meson1}
\end{eqnarray}
Here $\psi_2$ is the probability amplitude to find a quark and an antiquark
in the meson, $\psi_3$ is the probability amplitude to find a quark,
antiquark and a gluon in the meson etc. 

As in Sec. IV we evaluate the expression in Eq. (\ref{f2+}) explicitly.
The contribution from the first term (from the quark),
in terms of the amplitudes 
\begin{eqnarray}
	\sqrt{P^+}\psi_2(k_i^+, k_i^\perp) = && 
 		\Phi_2 (x_i, \kappa_i^\perp), \nonumber \\
	P^+\psi_3(k_i^+, k_i^\perp) = &&  \Phi_3(x_i, \kappa_i^\perp),
\end{eqnarray}
and so on, is
\begin{eqnarray}
	{F_2^q(x) \over x} = && \sum_{\sigma_1,\sigma_2} \int dx_2 \int
	d^2\kappa_1^\perp \int d^2 \kappa_2^\perp \delta (1 - x -x_2)
	\delta^2(\kappa_1 +  \kappa_2) \mid \Phi_2^{\sigma_1,
	\sigma_2}(x,\kappa_1^\perp; x_2 \kappa_2^\perp) \mid^2 \nonumber \\
&& + \sum_{\sigma_1, \sigma_2, \lambda_3} \int dx_2 \int dx_3 \int d^2
	\kappa_1^\perp \int d^2 \kappa_2^\perp \int d^2 \kappa_3^\perp 
	\delta(1 -x -x_2 -x_3) \delta^2(\kappa_1 + \kappa_2 + \kappa_3)
	\nonumber \\
&& ~~~~~~~~~~~~~~~~~~~~ \mid \Phi_3^{\sigma_1, \sigma_2, \lambda_3}(x,
\kappa_1^\perp; x_2, \kappa_2^\perp; x_3, \kappa_3^\perp) \mid^2 + ... ~~ .
\label{exact}
\end{eqnarray}
Again, the partonic interpretation of the $F_2$ structure function 
is manifest in this expression. Using different techniques and 
approximations, the same result has been also obtained by Brodsky 
and Lepage \cite{Brodsky}.

Contributions to the structure function from the second term in 
Eq. (\ref{f2+}) is
\begin{eqnarray}
	{F_2^{\bar q}(x) \over x} = && \sum_{\sigma_1,\sigma_2} \int dx_2 \int
	d^2\kappa_1^\perp \int d^2 \kappa_2^\perp \delta (1 - x -x_2)
	\delta^2(\kappa_1 +  \kappa_2) \mid \Phi_2^{\sigma_1,
	\sigma_2}(x_2,\kappa_2^\perp; x, \kappa_1^\perp) \mid^2 \nonumber \\
&& + \sum_{\sigma_1, \sigma_2, \lambda_3} \int dx_2 \int dx_3 \int d^2
	\kappa_1^\perp \int d^2 \kappa_2^\perp \int d^2 \kappa_3^\perp 
	\delta(1 -x -x_2 -x_3) \delta^2(\kappa_1 + \kappa_2 + \kappa_3)
\nonumber \\
&& ~~~~~~~~~~~~~~~~~~~~ \mid \Phi_3^{\sigma_1, \sigma_2, \lambda_3}(x_2,
\kappa_2^\perp; x, \kappa_1^\perp; x_3, \kappa_3^\perp) \mid^2 + ... ~~ .
\end{eqnarray}
The normalization condition guarantees that
\begin{eqnarray}
\int dx \big[ {F_2^{q}(x) \over x} + {F_2^{\bar q} (x) \over x} \big ] = 2
\end{eqnarray}
which reflects the fact that there are two valence particles in the meson.
Since the bilocal current component ${\bar {\cal J}}^+$ involves only
fermions explicitly, we appear to have missed the contributions from the
gluon constituents altogether. Gluonic contribution to the structure
function $F_2$ is most easily calculated by studying the hadron
expectation value of the conserved longitudinal momentum operator $P^+$. 

From the normalization condition, it is clear that the valence distribution
receives contribution from the amplitudes  $\Phi_2$, $\Phi_3$, ...  at any
scale $\mu$. This has interesting phenomenological implications. In the
model for the meson with only a quark-antiquark pair of equal mass, 
the valence distribution function will peak at $x = {1 \over 2}$. If there
are more than just the two particles in the system, the resulting valence
distribution will no longer be symmetric about $ x = {1 \over 2}$
 as a simple
consequence of longitudinal momentum conservation. 

The equation (\ref{exact}) as it stands is useful only when the bound state
solution in QCD is known in terms of the multi-parton wave functions. The
wave functions, as they stand,
span both the perturbative and non-perturbative sectors of the theory.
Great progress in the understanding of QCD in the high energy sector is made
in the past by separating the soft (non-perturbative) and hard
(perturbative) regions of QCD via the machinery of factorization. It is of
interest to see under what circumstances 
a factorization occurs in the formal result of Eq.
(\ref{exact}) and a perturbative picture of scaling violations emerges
finally. We shall explicitly address this issue in the following section.

\subsection{Perturbative picture of scaling violations in a bound state}

To address the issue of scaling violations in the structure function of 
the ``meson-like" bound state, it is convenient to separate the momentum
space into low-energy and high-energy sectors. Such a separation has been
introduced in the past in the study of renormalization  of bound
state equations \cite{Yukawa} in light-front field theory. The two sectors
are formally defined by introducing cutoff factors in the momentum space
integrals. How to cutoff the momentum integrals in a sensible and convenient
way in light-front theory is a subject under active research at the 
present time. Complications arise because of the possibility of large energy
divergences from both small $k^+$ and large $k^\perp$ regions.
In the following we investigate only the effects of logarithmic
divergences arising from
large transverse momenta, ignore subtleties arising from both small $x$
($x \rightarrow 0$)
and large $x$ ($x \rightarrow 1$) regions and subsequently use simple
transverse momentum cutoff. For complications arising from $x
\rightarrow 1$ region see Ref. \cite{Brodsky}. 
     
\subsubsection{Scale separation}
We define the soft region to be $\kappa^\perp < \mu$ and the hard
region to be $ \mu < \kappa^\perp < \Lambda$, where
$\mu$ serves as a factorization scale which separates soft and hard regions.
Since it is an intermediate scale introduced artificially purely for
convenience, physical structure function should be independent of $\mu$.
The multi-parton amplitude $\Phi_2$ is a function of a single relative
transverse momentum $\kappa_1^\perp$ and we define
\begin{eqnarray}
 \Phi_2 = \left \{ \begin{array}{c} \Phi_2^{s}, 
~~ 0 < \kappa_1^\perp < \mu, \\
     \Phi_2^{h}, ~~~ \mu < \kappa_1^\perp < \Lambda \end{array} \right.
\end{eqnarray}  
The amplitude $\Phi_3$ is a function of two relative momenta,
$\kappa_1^\perp$ and $\kappa_2^\perp$ and  we define
\begin{eqnarray}
	\Phi_3 = \left \{ \begin{array} {ll} 
	\Phi_3^{ss}, ~~~ & 0 < \kappa_1^\perp, \kappa_2^\perp < \mu \\
	\Phi_3^{sh}, ~ & 0 < \kappa_1^\perp < \mu, ~ \mu < \kappa_2^\perp < 
		\Lambda \\ 
	\Phi_3^{hs}, ~ & \mu < \kappa_1^\perp < \Lambda, ~ 0 <
		\kappa_2^\perp < \mu \\
	\Phi_3^{hh}, ~ & \mu < \kappa_1^\perp, \kappa_2^\perp <\Lambda 
		\end{array} \right.
\end{eqnarray}
Let us consider the quark distribution function $q(x) = {F_2(x) \over x}$
defined in Eq.(\ref{exact}). In presence of the ultraviolet cutoff
$\Lambda$, $q(x)$ depends on $\Lambda$ and schematically we have,
\begin{eqnarray}
	q(x,\Lambda^2) = \sum \int_0^{\Lambda} \Phi_2^2 + \sum 
		\int_0^\Lambda \int_0^\Lambda \Phi_3^2.
\end{eqnarray}
For convenience, we write,
\begin{eqnarray}
	q(x,\Lambda^2) =q_2(x,\Lambda^2) + q_3(x,\Lambda^2).
\end{eqnarray}
where the subscripts $2$ and $3$ denotes the two-particle and three-particle
contributions respectively. Thus, schematically we have,
\begin{eqnarray}
	q(x,\Lambda^2) = && q(x,\mu^2) + \sum \int_\mu^\Lambda \mid 
		\Phi_2^h \mid^2 \nonumber \\
	&&+ \sum \int_0^\mu \int_\mu^\Lambda \mid \Phi_3^{sh} \mid^2 + 
		\sum \int_\mu^\Lambda\int_0^\mu \mid \Phi_3^{hs} 
		\mid^2 \nonumber \\
	&& + \sum \int_\mu^\Lambda \int_\mu^\Lambda \mid \Phi_3^{hh} \mid^2.
\end{eqnarray}
We investigate the contributions from the amplitudes $\Phi_3^{sh}$ and
$\Phi_3^{hs}$ to order $\alpha_s$ in the following.       

\subsubsection{Dressing with one gluon}

We substitute the Fock expansion Eq. (\ref{meson1}) in 
Eq. (\ref{lfbe}) and
make projection with a three particle state $b^\dagger (k_1,\sigma_1)
d^\dagger(k_2, \sigma_2) a^\dagger(k_3, \sigma_3) \mid 0 \rangle $ from the
left. In terms of the amplitudes $\Phi_2$, $\Phi_3$, we get,
\begin{eqnarray}
\Phi_3^{\sigma_1 \sigma_2 \lambda_3}(x, \kappa_1; x_2, \kappa_2;
1-x-x_2, \kappa_3) = {\cal M}_1 + {\cal M}_2,
\end{eqnarray}
where the amplitudes
\begin{eqnarray}
	{\cal M}_1 = && { 1 \over E} (-) { g \over \sqrt{2 (2 \pi)^3}} T^a
		{ 1 \over \sqrt{1 - x - x_2}} ~V_1~ \Phi_2^{\sigma_1' 
		\sigma_2}(1-x_2, -\kappa_2^\perp; x_2,\kappa_2^\perp), 
\end{eqnarray}
and
\begin{eqnarray}
	{\cal M}_2 = && { 1 \over E} { g \over \sqrt{2 (2 \pi)^3}} T^a
		{ 1 \over \sqrt{1 - x - x_2}} ~V_2~ \Phi_2^{\sigma_1 
		\sigma_2'}(x,\kappa_1^\perp;1-x,-\kappa_1^\perp) 
\end{eqnarray}
with the energy denominator
\begin{eqnarray}
	E= \big[ M^2  - {m^2 + (\kappa_1^\perp)^2 \over x} -
		{m^2 + (\kappa_2^\perp)^2 \over x_2} - {(\kappa_3^\perp)^2 
		\over 1 - x -x_2} \big ],
\end{eqnarray}  
and the vertices     
\begin{eqnarray}
	V_1=\chi_{\sigma_1}^\dagger \sum_{\sigma_1'}\big [ { 2 
	\kappa_3^\perp \over 1 - x -x_2} - { (\sigma^\perp. \kappa_1^\perp
	- i m) \over x} \sigma^\perp + \sigma^\perp {(\sigma^\perp. 
	\kappa_2^\perp -im) \over 1-x_2} \big] \chi_{\sigma_1'}. 
		(\epsilon^\perp_{\lambda_1})^*,
\end{eqnarray}
and
\begin{eqnarray}
	V_2=\chi_{-\sigma_2}^\dagger \sum_{\sigma_2'}
		\big [ { 2 \kappa_3^\perp \over 1 - x -x_2} - \sigma^\perp
		{ (\sigma^\perp. \kappa_2^\perp
		- i m) \over x_2}  +  {(\sigma^\perp. \kappa_1^\perp -
		im) \over 1-x} \sigma^\perp 
		\big] \chi_{-\sigma_2'}. (\epsilon^\perp_{\lambda_1})^*
\end{eqnarray}

\subsubsection{Perturbative analysis}
For $\kappa_1^\perp$ hard and $\kappa_2^\perp$ soft,
$\kappa_1^\perp+\kappa_2^\perp \approx \kappa_1^\perp$ and the multiple
transverse momentum integral 
over $\Phi_3$ factorises into two independent integrals and the longitudinal
momentum fraction integrals become convolutions.
The contribution
from ${\cal M}_1$ to $\Phi_3$ is,
\begin{eqnarray}
\Phi_{3,1}^{\sigma_1,\sigma_2,\Lambda_3} (x,\kappa_1^\perp;x_2,\kappa_2^\perp;
1-x-x_2,-\kappa_2^\perp)= && - { g \over \sqrt{2 (2 \pi)^3}} T^a {x
\sqrt{1-x-x_2} \over 1 - x_2} {1 \over (\kappa_1^\perp)^2}
\nonumber \\ 
&& ~~ \chi^\dagger_{\sigma_1}\sum_{\sigma_{1}'}\big [ { 2 \kappa_1^\perp
\over 1-x-x_2}+ {\sigma^\perp. \kappa_1^\perp \over x} \sigma^\perp
\big] \chi_{\sigma_1}'
. (\epsilon^\perp_{\lambda_1})^*
\nonumber \\
&&~~~~~\Phi_2^{\sigma_1',\sigma_2}(1-x_2,-\kappa_2^\perp; x_2, \kappa_2^\perp).
\end{eqnarray}
Thus the contribution from ${\cal M}_1$ to the structure function is
\begin{eqnarray}
\sum \int \mid \Phi_{3,1}^{hs} \mid^2 = {\alpha_s \over 2 \pi} C_f
ln{\Lambda^2 \over \mu^2} \int_x^1 {dy \over y} P_{qq}({x \over y})
q_2(y,\mu^2), 
\end{eqnarray}
where
\begin{eqnarray}
P_{qq}\left({x \over y}\right) = { 1 + ({x \over y})^2 \over 1 - {x \over y}}.
\end{eqnarray}

For the configuration $\kappa_1^\perp$ hard, $\kappa_2^\perp$ soft,
contribution from ${\cal M}_2$ does not factorize and the asymptotic
behavior of the integrand critically depends on the asymptotic behavior of
the two-particle wave function  $\Phi_2$. To determine this behavior,
we have to analyze the bound state equation which shows that for large
transverse momentum $\Phi_2 (\kappa^\perp) \approx { 1 \over
(\kappa^\perp)^2}$. Thus contribution from ${\cal M}_2$ for scale evolution
is suppressed by the bound state wave function. Analysis of the interference
terms (between ${\cal M}_1$ and $ {\cal M}_2$) shows that their
contribution also is suppressed by the bound state wave function.

For the configuration $\kappa_1^\perp$ soft, $\kappa_2^\perp$ hard,
contributions from ${\cal M}_1$ and the interference terms are suppressed by
the wave function. Contribution from ${\cal M}_2$ factorises both in
transverse and longitudinal space and generate a pure wave function
renormalization contribution:   
\begin{eqnarray}
\sum \int \mid \Phi_{3,2}^{sh} \mid^2 = {\alpha_s \over 2 \pi} C_f 
ln{\Lambda^2 \over \mu^2} \int_0^1 dy {1 + y^2 \over 1-y}q_2(x,\mu^2).
\end{eqnarray} 

We have seen that even though the multi-parton contributions to the structure
function involve both coherent and incoherent phenomena, in the hard region
coherent effects are suppressed by the wave function.

\subsubsection{Corrections from normalization condition}
In the dressed quark calculation, we have seen that the singularity that
arises as $x
\rightarrow 1 $ from real gluon emission is canceled by the correction from
the normalization of the state (virtual gluon emission contribution
from wave function
renormalization). In the meson bound state calculation, so far we have
studied the effects of a hard real gluon emission. In this section we study
the corrections arising from the normalization condition of the quark
distribution in the composite bound state.

Collecting all the terms arising from the hard gluon emission contributing
to the quark
distribution function, we have,
\begin{eqnarray}
q(x,\Lambda^2) &&= q_2(x,\mu^2) + q_3(x,\mu^2) \nonumber \\
&& ~~~~~~ + { \alpha_s \over 2 \pi} C_f
ln{\Lambda^2 \over \mu^2} \int_x^1 {dy \over y} P_{qq} ({x \over y}) q_2(y,
\mu^2) \nonumber \\
&& ~~~~~~~ + {\alpha_s \over 2 \pi} C_f ln{\Lambda^2 \over \mu^2}
q_2(x,\mu^2) \int dy P(y).
\end{eqnarray}
We have a similar expression for the antiquark distribution function. 

The normalization condition on the quark distribution function should be
such that there is one valence quark in the bound state at any scale $Q$.
We choose the factorization scale $ \mu = Q_0$.
Let us first set the scale $ \Lambda = Q_0$. Then we have (in the truncated
Fock space)
\begin{eqnarray}
\int_0^1 dx ~ q_2(x,Q_0^2) + \int_0^1 dx ~ q_3(x ,Q_0^2) = 1.
\end{eqnarray}
Next set the scale $ \Lambda = Q$. We still require
\begin{eqnarray}
\int_0^1 dx ~ q_2(x,Q^2) + \int_0^1 dx ~ q_3(x,Q^2) = 1.
\end{eqnarray}
We note that the evolution of $q_3$ requires an extra hard gluon which is
not available in the truncated Fock space. Thus in the present approximation 
$ q_3(x,Q^2 ) = q_3(x,Q_0^2) $. 

Carrying out the integration explicitly, we arrive at 
\begin{eqnarray}
\int_0^1 dx ~ q_2(x,Q_0^2) \big[ 1 + {2 \alpha_s \over 2 \pi} C_f ln{Q^2 \over
Q_0^2} \int dy P(y)\big] + \int_0^1 dx ~ q_3(x,Q^2) = 1
\end{eqnarray}

Thus we face the necessity to $``$renormalize" our quark distribution
function. Let us define a renormalized quark distribution function 
\begin{eqnarray}
q_2^R (x,Q_0^2) = q_2(x, Q_0^2) \big [1 + 2 {\alpha_s \over 2 \pi} C_f ln{Q^2
\over Q_0^2} \int_0^1 dy P(y) \big]
\end{eqnarray}
so that, to order $\alpha_s$,
\begin{eqnarray}
\int_0^1 dx ~q_2^R(x,Q_0^2) + \int_0^1 dx ~ q_3(x,Q_0^2) = 1.
\end{eqnarray}
We have,
\begin{eqnarray}
q_2(x,Q_0^2) = q_2^R(x,Q_0^2)  \big [ 1 - 2 {\alpha_s \over 2 \pi} C_f
ln{Q^2 \over Q_0^2} \int_0^1 dy P(y) \big]. 
\end{eqnarray}
Collecting all the terms, to order $\alpha_s$, we have the normalized quark
distribution function,
\begin{eqnarray}
q(x,Q^2)&& = q_2^R(x,Q_0^2) \nonumber \\
&& ~~~~~~+ {\alpha_s \over 2 \pi} C_f ln{Q^2 \over Q_0^2} \int_0^1 dy
~ q_2^R(y, Q_0^2) \int_0^1 dz ~ \delta(zy-x)~ {\tilde P}(z) \nonumber \\
&& ~~~~~~ + q_3(x,Q^2) 
\end{eqnarray}
\
with $ {\tilde P}(z) = P(z) - \delta(z-1) \int_0^1 dy P(y)$.    
   
We see that just as in the dressed quark case, the singularity arising as $
x \rightarrow 1$ from real gluon emission is canceled in the quark
distribution function once the normalization
condition is properly taken in to account. From this derivation
we begin to recognize the emergence of the Altarelli-Parisi 
evolution equation.

\subsection{summary}
In this section we have carried out an analysis of the scale evolution of
structure functions of a meson-like composite system. We have separated the
parton transverse momenta into soft and hard parts. The three body wave
function which is a function of two relative momenta has soft, hard and
mixed components. The mixed components of the three body wave function which
are functions of soft and hard momenta are responsible for the scale
evolution of the soft part of the structure function in perturbation theory. 

In the analysis with wave functions, there are two contributions to the three
body wave function: One where the gluon is absorbed by the quark and
second where the gluon is absorbed by the anti-quark (spectator). There
appears a non-vanishing contribution  when the hard gluon is
absorbed by the anti-quark. This corresponds to the transition caused by the
interaction Hamiltonian when the active parton remains soft, while
a hard spectator makes transition to a soft spectator state. This leads to 
wave function renormalization of the spectator anti-quark but this is
eventually canceled by the normalization condition as discussed in detail
in Sec. VIB.4. This justifies a posteriori the prescription given in Sec. II that
we need to keep only those terms in $P^{-(H)}$ which cause transitions
involving the active parton. 

In the wave function analysis, there are also contributions that are omitted
a priori in the calculational scheme which lead to factorization in Sec. II.
All of these contributions are suppressed by the asymptotic behavior of the
bound state wave function as we have explicitly shown. In summary, the
detailed analysis carried out with the help of multi-parton wave functions
in Sec. VIB justifies the approximations made in Sec. II which lead to the
emergence of factorization to all orders in perturbation theory and to the
simple scale evolution picture.   

\section{Conclusion}
We have shown that a  
perturbative analysis in the light-front 
Hamiltonian
formalism  leads to the factorization scheme proposed in Ref. \cite{paper1}.
It is shown that
the scaling violations due to perturbative QCD
corrections can be rather easily addressed in this framework by simply
replacing the hadron target by dressed parton target and then carrying 
out a systematic expansion in the coupling constant $\alpha_s$ based on 
the perturbative QCD expansion of the dressed parton target.
The calculational procedure utilizes techniques of old-fashioned
perturbation theory main ingredients of which are transition matrix
elements and energy denominators.

The main advantage of the present method can be summarized as follows. The
bilocal currents are defined in the light-front gauge $A^+=0$ and since the
bilocality is only in the light-front longitudinal ($x^-$) direction, the
path-ordered exponential between fermion field operators in the bilocal
current is replaced by
unity ({\it irrespective} of which component of the current is considered).
This results in an extremely simplified operator structure and a
straightforward parton picture. Further, the calculations do not employ
Feynman propagators and as a result we encounter neither the usual problems
associated with using a non-covariant gauge in a covariant calculation nor
the problems associated with the unphysical pole of the propagator. The
calculations are straightforward and $\gamma_5$ or presence of quark masses
pose no special problem. The physical picture is very clear at every 
stage of the calculation. Also the regularization scheme 
used in this framework for perturbative contribution
can be directly applied to the construction of hadronic bound states
which is the major topic of current research on light-front field theory
\cite{Wilson94,review}. Thus, once the light-front bound state structure are 
found, a complete theoretical understanding of structure functions 
can become possible.

In addition, the approach uses probability amplitudes rather than 
probability densities
and hence interference effects are easy to handle. Exploiting this feature,
we have clarified the parton interpretation of the matrix elements of the
transverse component of biocal vector and axial vector currents. 
We have presented real and virtual corrections to the structure functions
$F_2$ and $g_1$ for a dressed quark and gluon in a transparent manner. 
The splitting functions are extracted and the longitudinal momentum and
helicity sum rules are verified explicitly to order $\alpha_s$. 

We have carried out, with the help of multi-parton wave functions,
a detailed analysis of the scale evolution of the structure function of a
composite system which
 justifies the approximations made in Sec. II which lead to the
emergence of factorization to all orders in perturbation theory and to the
simple scale evolution picture.   
A complete fourth order calculation 
is necessary to establish the viability of the new approach for the
perturbative domain. Such a calculation is presently under way.  Also, 
the main contribution to DIS structure functions, the nonperturbative
QCD dynamics, is also in progress. We shall leave the  discussion of these
topics for future publications.
      
\acknowledgments
We acknowledge useful discussions with Stan Brodsky amd James Vary.
This work is partially supported by NSC86-2816-M001-009R-L, 
NSC86-2112-M001-020 and Physics Institute of Academia Sinica(WMZ). 

\end{document}